\documentclass[conference,usenatbib]{basi}

\usepackage[T1]{fontenc}
\usepackage[british]{babel}
\usepackage[varg]{txfonts}

\usepackage{rotating}
\usepackage{dcolumn}

\begin{document}
\vspace{-0.4cm}
\title[Multi-band variability of blazars]{Optical-NIR variability of blazars on diverse timescales}
\vspace{-0.6cm}
\author[A.~Agarwal et~al.]%
       {A.~Agarwal$^1$\thanks{email: \texttt{aditi@aries.res.in}},
       and A.~C.~Gupta$^{1}$\\
       $^1$Aryabhatta Research Institute of Observational Sciences (ARIES),
Manora Peak, Nainital -- 263002, India}

\pubyear{2015}
\volume{12}
\pagerange{\pageref{firstpage}--\pageref{lastpage}}
%\status{submitted}
\vspace{-0.6cm}
\date{Received --- ; accepted ---}

\maketitle

\label{firstpage}

\vspace{-0.5cm}

\begin{abstract}

To search for optical variability on a wide range of
timescales, we have carried out photometric
monitoring of 3C 454.3, 3C 279 and S5 0716+714. CCD
magnitudes in B, V, R and I pass-bands were
determined for $\sim$ 7000 new optical observations from
114 nights made during 2011 – 2014, with an average
length of $\sim$ 4 h each, at seven optical telescopes. We
measured multiband optical flux and colour variations
on diverse timescales. We also investigated its
spectral energy distribution using B, V, R, I, J and K
pass-band data. We discuss possible physical causes of the observed spectral variability.

\end{abstract}
\vspace{-0.5cm}
\vspace{-0.5cm}

\section{Introduction}\label{s:intro}
\vspace{-0.5cm}
Some of the brightest radio-loud Active Galactic Nuclei (AGNs), called blazars, are understood to
have relativistic jets viewed at an angle of $\leq$ 10$^{\circ}$ from the line of sight (LOS) 
(e.g., Urry \& Padovani 1995).
BL Lacs (Featureless optical spectra) and FSRQs
(prominent optical emission lines ) together form blazars class.
Variability timescales in blazars are divided into three classes: flux variations
over a timescale of few minutes to less than a day are called as intra-day variability (IDV; Wagner \& Witzel 1995);
those on timescale of days to few months 
are as short time variability (STV);
while the changes from several months to many years are usually called long term variability.
The key motivation of this work was to search for optical flux and colour variability of 3C 454.3, S5 0716+714 and
3C 279 on diverse
time scales, including analyses of colour-magnitude variations, inter-band cross correlations,
and optical/NIR 
SEDs.
% \begin{figure}
%
%\centerline{\includegraphics[width=3cm]{fig1a.eps} \qquad
%            \includegraphics[width=3cm]{fig3h.eps}}
%
%\caption{This figure illustrates how to multiple plots, both horizontally or
%vertically. It also shows how to rotate figures by $\pm 90^\circ$, keeping them
%aligned vertically.\label{f:many}}
%
%\end{figure}
 
%------------------------------------------------------------------------------%

\vspace{-0.5cm}
\section{Observations and Analysis}
\vspace{-0.4cm}
Our optical photometric observations of the blazars were performed in the B, V, R, and I pass-bands,
using seven telescopes, two in India, one in Greece, and four in Bulgaria, all equipped with CCD detectors. 
The preliminary processing of the raw data was
carried through standard procedures in IRAF
software. We performed photometry of the
data to find the instrumental magnitudes of the
blazars and the comparison stars by concentric
circular aperture photometric technique with
the DAOPHOT II software.
The IDV of blazars was examined employing both the popularly used C-statistic and the more reliable F-test (Agarwal \& Gupta 2015).
The percentage variation on a given night is calculated by using the variability amplitude parameter $A$,
introduced by Heidt \& Wagner (1996),

\vspace{-0.5cm}
\section{Results}
\vspace{-0.4cm}
During our 37 nights of monitoring for IDV we
found flux variations, on intraday
time-scales for 21 nights in B, V, R or I with amplitudes of variability up to
31\% for 3C 454.3, 9.2\% for 3C 279 and
13.5\% for S5 0716+714.
For each night, we calculated a linear fit of the
colour indices, CI, against V magnitude:
CI=mV + c.
The linear Pearson correlation coefficient, r,
and the corresponding null hypothesis
probability, p, also calculated.
A positive slope indicates a positive
correlation between the colour indices and
apparent magnitude of the blazar which
implies the general trend of BWB.
BWB trend was dominant
during our observations. Such flattening
may be interpreted as jet dominated
synchrotron emission, where increasing flux is
related to a hardening of non-thermal electron
spectrum, possibly indicating an enhanced
particle-acceleration efficiency (Diltz \& B\"ottcher 2014).

\vspace{-0.6cm}
\begin{figure}
\centerline{\includegraphics[width=9cm]{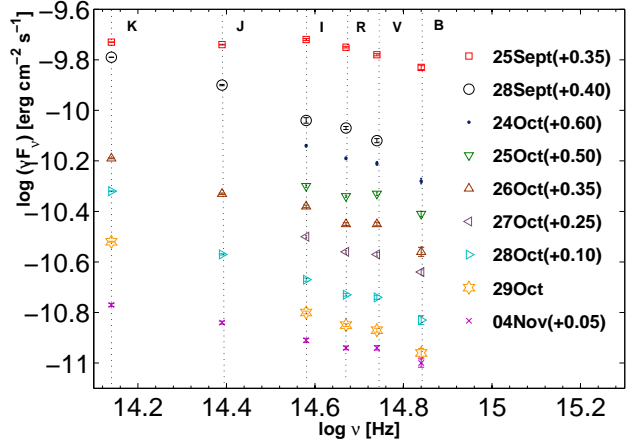}}
\caption{SED results for 3C 454.3 in near-IR-optical frequency range.}
\end{figure}

\end{document}